\pdfoutput=1   
\documentclass{article}

\usepackage[T1]{fontenc}
\usepackage{microtype}
\usepackage{graphicx}
\usepackage{booktabs}
\usepackage{amsmath}
\usepackage{amssymb}
\usepackage{multirow}
\usepackage{xcolor}
\usepackage{tikz}
\usetikzlibrary{positioning,arrows.meta,fit}
\usepackage{hyperref}

\usepackage[accepted]{mlsys2025}
\makeatletter
\renewcommand{\Notice@String}{Preprint.}
\makeatother

\newcommand{\sys}{TableTune}
\newcommand{\term}[1]{\emph{#1}}

\mlsystitlerunning{Measured, Self-Correcting Draft-Length Tables for Dynamic Speculative Decoding}

\begin{document}

\twocolumn[
\mlsystitle{\sys{}: Measured, Self-Correcting Draft-Length Tables\texorpdfstring{\\}{ }for Dynamic Speculative Decoding}

\begin{mlsysauthorlist}
\mlsysauthor{Abhishek Malvankar}{ibm}
\mlsysauthor{Paul Castro}{ibm}
\mlsysauthor{Nick Mitchell}{ibm}
\end{mlsysauthorlist}
\mlsysaffiliation{ibm}{IBM Research}
\mlsyscorrespondingauthor{Abhishek Malvankar}{asmalvan@us.ibm.com}
\mlsyscorrespondingauthor{Paul Castro}{castrop@us.ibm.com}
\mlsyscorrespondingauthor{Nick Mitchell}{nickm@us.ibm.com}

\mlsyskeywords{Speculative Decoding, LLM Serving, vLLM, llm-d, Performance Modeling}

\vskip 0.3in

\begin{abstract}
Speculative decoding speeds up large language model (LLM) serving only while the GPU has spare compute, so the best \emph{draft length}, the number of tokens a small drafter guesses per step, shrinks as the batch grows. vLLM lets operators set the draft length for each batch size with a hand-written table, but nothing measures or corrects that table. We present \sys{}. Its \emph{tuner} times the decoding step for every batch size and draft length on vLLM's own serving path and computes, in minutes, the table that maximizes throughput. Its \emph{controller}, a vLLM plugin, tracks how many guesses are accepted in live traffic and rebuilds the table while serving. We measure \sys{} only against other ways to speculate. On an H100, the tuned table delivers at least 94\% of the throughput of the best fixed draft length at every load in four of five drafter settings (88\% in the fifth) without the operator having to know which length that is. Against reimplementations of TurboSpec- and Nightjar-style control on the same engine, its worst-case loss is never higher than theirs, and with the EAGLE-3 drafter it is 3\% against 9\% and 12\%. On a mixture-of-experts model, vLLM's documented table loses up to 18\% and the tuned table at most 3\%. Starting from mistuned tables, the controller raises throughput by up to 58\% (up to 45\% on vLLM's default runner), and after a workload switch with the DFlash drafter it recovers 96\% of a stale table's loss. In llm-d's optimized-baseline deployment as shipped, two Qwen3-32B replicas on four GPUs, the tuned table is up to 32\% faster than a fixed draft length of 3 and up to 25\% faster than vLLM's documented table at high load, and the controller raises the documented table's throughput by up to 29\%.
\end{abstract}
]

\printAffiliationsAndNotice{}
\raggedbottom

\section{Introduction}
\label{sec:intro}

Generating text with an LLM is slow because it is sequential: each new token requires a full pass through the model, and each pass mostly waits on reading the model's weights from GPU memory~\cite{pope2023efficiently}. \term{Speculative decoding}~\cite{leviathan2023fast,chen2023accelerating} hides part of this cost. A small, fast \term{drafter} guesses the next few tokens; the \term{target model} checks all guesses in one pass, keeps those it agrees with, and adds one token of its own. Because every kept token is verified, the output is exactly what the target would have produced alone.

Speculation is not free. Every guess is extra work, and wrong guesses are thrown away. On an idle GPU the extra work is nearly free, because the GPU waits on memory anyway; on a busy GPU serving many requests it slows every step, and speculation can make serving slower than not speculating~\cite{liu2024turbospec}. The best number of guesses per step, the \term{draft length}, therefore depends on load.

vLLM~\cite{kwon2023efficient} recently added \term{Dynamic Speculative Decoding} (Dynamic SD)~\cite{vllm2026dynamicsd}, which lets the draft length depend on the \term{batch size}, the number of requests in a step. The operator supplies a table such as ``batch sizes 1--64: guess 3 tokens; 65--128: guess 1; above: none.'' The table is written by hand, often copied from the documentation, and nothing checks or corrects it. Users report throughput collapsing when the batch oscillates around a table edge~\cite{vllm2026issue49548} and Dynamic SD running 21--31\% slower than no speculation on an H100~\cite{vllm2026issue49986}.

\begin{figure}[!b]
\centering
\resizebox{\columnwidth}{!}{\begin{tikzpicture}[
  font=\footnotesize, >={Stealth[length=1.7mm]},
  box/.style={draw=black!55, rounded corners=1.5pt, fill=#1, align=center, inner sep=2.5pt, text width=29mm, minimum height=9mm},
  box/.default=white,
  lab/.style={font=\scriptsize\itshape, text=black!60},
  small/.style={font=\scriptsize},
  flow/.style={->, draw=black!70, thick},
  live/.style={->, draw=blue!60!black, thick}]
\node[box=black!4] (cost) at (0,0) {step-cost profiler\\{\scriptsize serving path, forced rejection,}\\[-0.4ex]{\scriptsize grid of $B \times k$}};
\node[box=black!4] (acc) at (3.6,0) {acceptance profiler\\{\scriptsize 160 sample prompts,}\\[-0.4ex]{\scriptsize serving temperature}};
\node[box=black!4, text width=40mm] (build) at (1.8,-1.75) {table builder: $\arg\max_k\; B\,\tau(k)/T(B,k)$\\[-0.2ex]{\scriptsize per batch size, then tie and width rules}};
\draw[flow] (cost.south) -- node[left, lab]{$T(B,k)$} (cost.south |- build.north);
\draw[flow] (acc.south) -- node[right, lab]{$a_i$} (acc.south |- build.north);
\node[fit=(cost)(acc)(build), draw=black!35, dashed, rounded corners, inner sep=2.2mm] (tuner) {};
\node[lab, anchor=south west] at (tuner.north west) {tuner: offline, once per GPU, engine, model, drafter};
\node[box] (sched) at (0,-4.35) {scheduler\\{\scriptsize live batch $B \rightarrow$ draft length $k$}};
\node[box] (exec) at (3.6,-4.35) {decode step\\{\scriptsize draft $k$ tokens, verify}};
\node[box=blue!6, text width=65mm, minimum height=10mm] (ctrl) at (1.8,-6.0) {controller plugin: re-estimate $\hat a_i$ from live counts and the prior\\[-0.2ex]{\scriptsize rebuild every $N$ steps with $T(B,k)$; hold near edges ($h$); exploration, prior decay}};
\draw[flow] (sched) -- node[above, lab]{$k$} (exec);
\draw[live] (exec.south) -- node[right, lab, text=blue!50!black]{accepted, drafted} (exec.south |- ctrl.north);
\draw[live] (sched.south |- ctrl.north) -- node[left, lab, text=blue!50!black]{new table} (sched.south);
\node[fit=(sched)(exec)(ctrl), draw=black!35, dashed, rounded corners, inner sep=2.2mm] (vllm) {};
\node[lab, anchor=north west] at (vllm.south west) {vLLM serving (no engine changes)};
\draw[flow] (build.south) -- ++(0,-0.75) node[pos=0.55, right, lab]{table, vLLM format} -| (sched.north);
\end{tikzpicture}}
\vspace{-5mm}
\caption{\sys{}. The tuner measures step time on vLLM's serving path and acceptance on sample prompts, and writes a table in vLLM's own format. While serving, the controller plugin reads vLLM's acceptance counters, re-estimates acceptance and rebuilds the table with the tuner's step times.}
\label{fig:design}
\vspace{-3mm}
\end{figure}
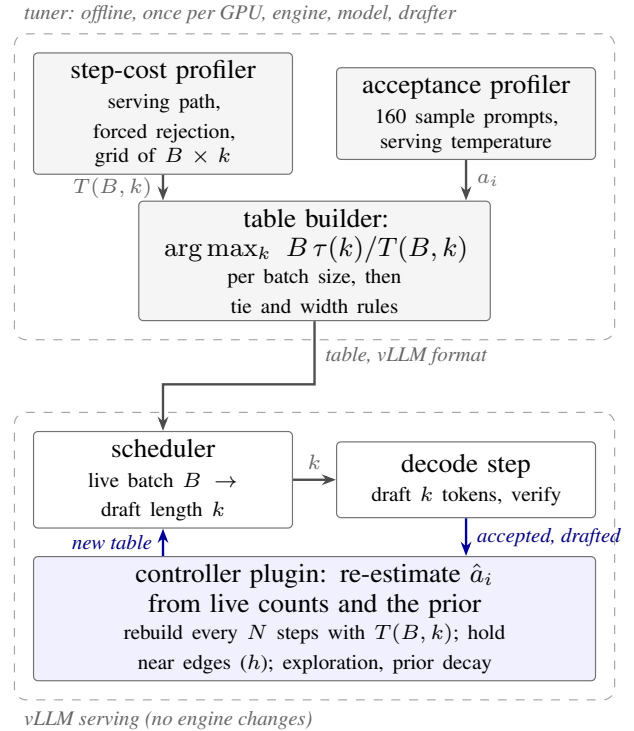

\paragraph{\sys{}.} \sys{} keeps vLLM's mechanism and makes its table measured and self-correcting (Figure~\ref{fig:design}). Its \textbf{tuner} runs once per GPU, engine, model and drafter: it times one decoding step for a grid of batch sizes and every draft length through vLLM's own serving path, measures how often guesses are accepted, and computes the table that maximizes useful tokens per second at every batch size. Its \textbf{controller}, a vLLM plugin that needs no engine changes, re-estimates acceptance from live traffic and rebuilds the table while serving. Both emit vLLM's own table format. Prior work established the ingredients: TurboSpec~\cite{liu2024turbospec} chooses the draft length that maximizes goodput from profiled step times, \citet{su2023synergy} make it depend on the batch size, and Nightjar~\cite{li2025nightjar} learns it online. TurboSpec and Nightjar change the engine's decisions; \sys{} instead makes vLLM's existing production table measured and self-correcting, with step times measured on the serving path rather than fitted, a controller that runs as a plugin, and a three-GPU study of what that table can deliver and what it needs from the engine.

\paragraph{Contributions.}
\begin{enumerate}\setlength{\itemsep}{1pt}
\item \emph{A tuner that measures the table} (Sections~\ref{sec:tuner}, \ref{sec:eval-tables}). On an H100 with vLLM's V2 runner, which keeps full CUDA graphs for every draft length (Section~\ref{sec:background}), at every load, the tuned table delivers at least 94\% of the throughput of the best fixed draft length for that load with the EAGLE-3 drafter (code prompts, chat prompts, a 4-bit target) and with n-gram drafting, and at least 88\% with DFlash, where any single fixed draft length drops to 81\% or less at some load. Under \emph{jumping load}, which changes every 50 seconds, it delivers in every 50-second window at least 98.5\% of the throughput of the fastest alternative (fixed draft length 1 or 2, or vLLM's documented table). On a mixture-of-experts target the best length stays long at every load, and the tuned table stays within 2.9\% where vLLM's documented table, written with dense models in mind, loses up to 17.5\%. On a T4 and an A10G, for full-precision targets, the tables deliver at least 94\% of it, where every fixed draft length drops 7.6--43.6\% below it at some load.
\item \emph{A comparison with prior adaptive control} (Section~\ref{sec:eval-adaptive}): on the same engine and lengths, reimplementations of TurboSpec-style per-step control lose up to 9.3\% with EAGLE-3, because a fitted step-time model misses a jump in step time, and of Nightjar-style bandits up to 11.8\%, because every batch size keeps exploring; the tuned table loses at most 3.3\%.
\item \emph{A controller that repairs tables while serving} (Sections~\ref{sec:controller}, \ref{sec:eval-ctrl}): up to +58\% on the H100 from a mistuned table, +23\% for DFlash started from another drafter's table, and 96\% of a stale table's 10.8\% loss after a workload switch with DFlash on the H100 (72--85\% on the A10G).
\item \emph{End-to-end gains in llm-d} (Section~\ref{sec:eval-llmd}), a Kubernetes-native distributed serving stack built on vLLM~\cite{llmd}: on its optimized-baseline guide as shipped (Qwen3-32B with tensor parallelism over two GPUs, TP=2), without changes to llm-d or vLLM, up to +32\% throughput over fixed length 3 and +25\% over vLLM's documented table at high load, and up to +29\% from the controller started from that table; with four Qwen3-8B replicas, up to +22\% and +40\%.
\item \emph{What table-based speculation needs from the engine} (Section~\ref{sec:eval-engine}): full CUDA graphs per draft length (vLLM's default V1 runner loses 30--43\% on the H100 with any table, regardless of host cores), a real off switch, and length switches that do not stall the engine.
\end{enumerate}

\section{Background}
\label{sec:background}

\paragraph{Batches and decoding.} A serving engine runs one forward pass over a \term{batch} of requests per \term{step}; the \term{batch size} $B$ is the number of requests in it. With \term{continuous batching}~\cite{yu2022orca} requests join and leave every step, so $B$ follows the traffic. Our benchmarks set the \term{concurrency} $c$, the number of open requests. The live batch, vLLM's count of running requests, is smaller: on average about half to two thirds of $c$ in our closed-loop runs. Tables are indexed by the live batch; our figures plot $c$. In a decode step the GPU reads every weight once, so with few tokens per step the step time is set by memory bandwidth and extra tokens cost little, up to the \term{ridge point}~\cite{williams2009roofline} (about 200 tokens per step on a T4, 210 on an A10G and 300 on an H100 SXM).

\paragraph{Speculative decoding.} A drafter proposes $k$ tokens. The target scores them in one pass and accepts guesses up to the first disagreement, then appends one token of its own~\cite{leviathan2023fast,chen2023accelerating}. With $a_i$ the probability that guess $i$ and all before it are accepted, the expected \term{tokens per step} is $\tau(k) = 1 + \sum_{i=1}^{k} a_i$. EAGLE-style drafters~\cite{li2024eagle,li2024eagle2,li2025eagle3} are small heads on the target's features; a \term{draft model} is a separate small LLM; \term{n-gram} drafting~\cite{saxena2023prompt} copies prompt text; DFlash~\cite{chen2026dflash} drafts a block of tokens in one pass of a small diffusion model.

\paragraph{Goodput.} We maximize \term{goodput}, kept tokens per second~\cite{liu2024turbospec}: the tuner predicts it with Equation~\ref{eq:goodput}, and we measure it as output tokens per second (throughput). With step time $T(B,k)$,
\begin{equation}
G(B,k) = B \cdot \tau(k) \,/\, T(B,k) .
\label{eq:goodput}
\end{equation}
Speculation raises the numerator and the denominator; whether it pays depends on which grows faster at that $B$.

\paragraph{vLLM Dynamic SD and CUDA graphs.} vLLM takes a table \texttt{num\_speculative\_tokens\_per\_batch\_size} of ranges $[B_{\text{start}}, B_{\text{end}}, k]$ and looks up $k$ for each step~\cite{vllm2026dynamicsd}; the documentation's example is $[[1,64,3],[65,128,1],[129,512,0]]$. A \term{CUDA graph}~\cite{cudagraphs} replays a recorded sequence of kernels with one launch, saving CPU launch time. With a table, vLLM's default V1 model runner replaces its \emph{full} decode graphs with \emph{piecewise} graphs, because the tokens per request now vary; its V2 runner keeps full graphs per draft length since October 2026~\cite{vllm2026pr57053}.

\section{Motivation}
\label{sec:motivation}

\begin{figure}[t]
\centering
\includegraphics[width=\columnwidth]{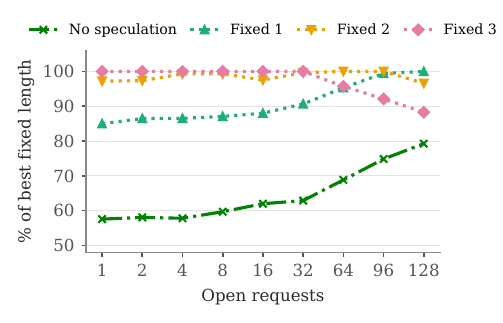}
\vspace{-6mm}
\caption{No fixed draft length is best at every load. H100 (V2 runner), Qwen3-4B + EAGLE-3, code prompts at temperature 1.0, mean of three runs: throughput as \% of the best fixed draft length at each load.}
\label{fig:h100}
\vspace{-3mm}
\end{figure}

Figure~\ref{fig:h100} shows the problem on an H100 with Qwen3-4B~\cite{qwen3} and its EAGLE-3 drafter~\cite{li2025eagle3}. \emph{The best length changes with load:} length 3 is best at low load and length 1 at 128 requests; fixed length 1 loses 15\% at 1 request and length 3 loses 12\% at 128. \emph{The best table changes with the GPU and the engine:} the tuned tables are $[[1,18,3],[19,43,2],[44,128,1]]$ on a T4, $[[1,16,3],[17,60,2],[61,91,1],[92,128,0]]$ on an A10G and $[[1,50,3],[51,128,2]]$ on the H100 with the V2 runner; the H100's plain step grows only from 4.4 to 6.1~ms between a batch of 1 and 128, the T4's from 35.7 to 92.3~ms. \emph{The engine decides whether any table can pay:} on the default V1 runner a table costs about 40\% of the throughput at low load (Table~\ref{tab:main}). \emph{Acceptance changes with the traffic:} EAGLE-3's tokens per step at length 3 range from 1.81 (translation) to 2.43 (math) across seven Spec-Bench~\cite{xia2024unlocking} and code workloads, on every GPU we measured.

These observations suggest a split. Step time depends on hardware, engine and batch size, is stable, and can be measured once offline. Acceptance depends on traffic, drifts, and should be tracked online. \sys{} follows this split.

\section{Design}
\label{sec:design}

\subsection{The tuner}
\label{sec:tuner}

\paragraph{Step time.} Let $K$ be the longest draft length the server may use, vLLM's \texttt{num\_speculative\_tokens} (3 for EAGLE-3, 5 for n-gram and the draft model, 8 for DFlash). For each draft length $k$ from 0 to $K$ the tuner starts vLLM with the real drafter and a one-range table $[[1,B_{\max},k]]$, so the timed path is exactly the serving path, including the graph mode the server will use. For each batch size $B$ it sends $B$ prompts of the workload's typical length (32 tokens by default) twice, asking once for 1 output token and once for 65; the difference is 64 decode steps, so $T(B,k) = (t_{65}-t_{1})/64$. It keeps the median of three repeats, over every batch size up to 16 and then in steps of 4 to 8 up to $B_{\max}$. To keep the batch fixed while timing, it uses vLLM's test mode that rejects every guess: each step does the full drafting and verification work of length $k$, but all $B$ requests advance one token per step and finish together. The benefit of accepted guesses is added back through $\tau(k)$. Work that grows with the number of accepted tokens is not timed this way; Section~\ref{sec:eval-cost} measures what that misses.

\paragraph{Acceptance.} The tuner then measures how often guesses are accepted. It runs the drafter at the longest length, $K$ guesses per step, on 160 sample prompts, at the temperature the server will use, and reads vLLM's own published speculative-decoding metrics, \texttt{vllm:spec\_decode\_}\allowbreak\texttt{num\_drafts} and \texttt{vllm:spec\_decode\_}\allowbreak\texttt{num\_accepted\_}\allowbreak\texttt{tokens\_per\_pos}: how many drafts were made and, for every position $i$, how many of them kept guess $i$. Their ratio is $a_i$. One run covers every shorter length too, because a draft of length $k$ is just the first $k$ guesses of a draft of length $K$.

\paragraph{The table.} For each batch size $B$ the tuner computes the goodput of every draft length (Equation~\ref{eq:goodput}) and picks the highest. Two rules keep the table short. It keeps the previous batch size's length unless another is more than 0.5\% better, so near-ties do not create edges, and it merges any range narrower than 25\% of its first batch size into a neighbour. Some drafters guess only on some steps: n-gram drafts only when the text it has seen contains a match. For them the tuner multiplies acceptance by the share of steps with a guess, the \term{proposal rate}, but still charges the full step time $T(B,k)$, since a step in which some requests draft costs about as much as one in which all do. From the same profiles, a GPU-free \emph{screening} predicts each fixed length's worst loss, so an operator can keep a fixed length when a table would not help.

\subsection{The controller}
\label{sec:controller}

The controller is a vLLM general plugin that answers the scheduler's draft-length lookup. It learns acceptance from live traffic and turns it into a table.

\paragraph{What it observes.} The controller adds no instrumentation of its own. After every step it receives the counts that vLLM's scheduler records for those same published metrics: for each request, how many guesses were drafted ($d$) and how many were accepted ($m$; verification stops at the first wrong guess, so the accepted guesses are always the first $m$). The controller keeps two counts per draft position $i$: $\text{drafted}_i$, the number of drafts that reached position $i$ ($d \ge i$), and $\text{accepted}_i$, the number in which guess $i$ was accepted ($m \ge i$). A request that drafts 3 guesses and keeps 2 adds one to $\text{drafted}_{1..3}$ and to $\text{accepted}_{1..2}$. The ratio $\text{accepted}_i/\text{drafted}_i$ estimates $a_i$, the probability that the first $i$ guesses are all accepted.

\paragraph{From acceptance to draft length.} Acceptance sets what each extra guess is worth: since $\tau(k) = 1 + \sum_{i\le k} a_i$, going from length $k-1$ to $k$ gains $a_k$ tokens per step and costs $T(B,k) - T(B,k-1)$. The gain falls with $k$ because $a_k$ falls, and the cost grows with $B$, so for each $B$ the best length is where the next guess stops paying, the maximum of Equation~\ref{eq:goodput}. Higher acceptance therefore keeps long drafts up to larger batches; lower acceptance moves the table's edges to smaller batches and shorter lengths, down to length 0. The step times never change while serving, so acceptance is the only input the controller updates.

\paragraph{Estimating and rebuilding.} Every $N$ steps it computes $\hat a_i = (w\, a_i^{\text{prior}} + \text{accepted}_i)/(w + \text{drafted}_i)$, a blend of the tuner's values (worth $w=2000$ observations) and the live counts, which fade by half after each rebuild so recent traffic dominates, and rebuilds the table from Equation~\ref{eq:goodput} with the tuner's step times. A position that is never drafted keeps its prior: while the table runs length 1 or 0 the controller learns nothing about later guesses. Two options matter when the traffic changes: \emph{exploration} (one step at the longest allowed length every 50 steps, so every position keeps being observed) and \emph{prior decay} (the prior's weight also halves at each rebuild). Near a table edge it holds the current length while the batch size is within $h$ of the edge (\term{hysteresis}), avoiding the oscillation of~\cite{vllm2026issue49548}. By default it uses only the lengths in its starting table, which have profiles and captured graph sizes; under V1 an opt-in mode allows any profiled length, while under V2 only the starting table's lengths have full CUDA graphs.

\section{Implementation}
\label{sec:impl}

\sys{} is about 1{,}880 lines of Python with 104 test functions. The tuner is one command (\texttt{dsd-tuner run}). Because the table is a vLLM argument and the controller an installed plugin enabled by environment variables, \sys{} needs no engine changes and deploys into serving stacks such as llm-d by configuration alone. The tuner runs unchanged on vLLM's V1 and V2 runners. Code, per-GPU tables, benchmark scripts, raw results and analysis scripts are available at \url{https://github.com/asm582/tabletune}.

\section{Evaluation}
\label{sec:eval}

We answer six questions. \emph{Q1} (Section~\ref{sec:eval-tables}): does a measured table hold up against every fixed draft length, across loads, drafters and GPUs? \emph{Q2} (Section~\ref{sec:eval-adaptive}): how does it compare with prior adaptive controllers? \emph{Q3} (Section~\ref{sec:eval-ctrl}): does the controller repair mistuned tables and follow a workload change? \emph{Q4} (Section~\ref{sec:eval-llmd}): do the gains carry over to a production stack like llm-d? \emph{Q5} (Section~\ref{sec:eval-engine}): what must the engine provide? \emph{Q6} (Section~\ref{sec:eval-cost}): what does \sys{} cost, and how reliably can step times be measured?

\subsection{Setup}
\label{sec:setup}

\paragraph{Hardware and software.} Our main platform is an H100 80~GB SXM (3.35~TB/s, BF16) on an OpenShift cluster, a whole GPU with a 16-CPU quota on a Xeon 8480+ host. We ran the main matrix of tuner, sweep and controller experiments twice on it: with vLLM \texttt{353e491ef} and its default V1 runner, and with a vLLM nightly (\texttt{43b4aaea}, October 2026) and the V2 runner, which keeps full CUDA graphs per draft length. Because a table cannot pay under V1 (Section~\ref{sec:eval-engine}), all later studies use V2 only. The repository lists every study with its scripts, tables and raw results. For comparison we ran the same experiments, with the same scripts, prompts and loads and vLLM's V1 runner, on a Tesla T4 (FP16) and an A10G (BF16), both AWS instances with 2-core hosts. Targets are Qwen3-4B, its 4-bit AWQ version~\cite{lin2024awq}, the mixture-of-experts Qwen3-30B-A3B~\cite{qwen3} and, in llm-d, Qwen3-32B at TP=2 and Qwen3-8B; drafters are EAGLE-3, n-gram, Qwen3-0.6B as a draft model, and DFlash.

\paragraph{Workloads and load.} AIPerf~\cite{aiperf} keeps $c$ requests open in a closed loop; a \term{sweep} measures every load $c$ from 1 to 128, with $4c$ requests per level after $c$ warm-up requests, and the maximum batch is 128. The main workload is \emph{code}: InstructCoder~\cite{li2024instructcoder} instructions (about 32 prompt tokens), answers up to 256 tokens, temperature 1.0. n-gram uses \emph{code-apply} (an instruction plus the code to edit, about 151 tokens in, answers up to 512 tokens, greedy); the chat study uses Spec-Bench (greedy, up to 64 requests).

\paragraph{Isolation and statistics.} Each H100 V1 study ran alone on an otherwise idle node, because piecewise CUDA graphs turned out to be sensitive to host contention (Section~\ref{sec:eval-engine}); the V2 studies ran on nodes shared with other teams' GPU jobs, which full graphs are insensitive to ($\pm$4\%). We report output tokens per second per GPU, the mean of three runs unless marked; the run-to-run spread on the H100 is 1.4\% (median) and 5.7\% (95th percentile) with V2 and 2.5\%/11.2\% with V1, and at most 4.1\% and 6.5\% on the T4 and A10G. We treat gaps below about 2\% as ties. Unless stated otherwise the controller uses hysteresis $h=8$ and rebuilds the table every 200 steps; the workload-switch runs use its defaults ($h=4$, every 500 steps) with exploration every 50 steps and prior decay 0.5. Draft lengths and acceptance reported in the evaluation come from the same vLLM metrics, read from each server's \texttt{/metrics} endpoint, so every number can be checked against vLLM's own accounting.

\paragraph{Baselines.} \emph{Best fixed} at a load is the fastest fixed draft length there, without a table and with full graphs, chosen after the fact; losses are against it. \emph{Best single fixed length} is the one length with the smallest worst-case loss over all loads, also chosen after the fact: an operator would have to benchmark every length to find it. We credit \sys{} only with differences from such speculative configurations (and from vLLM's documented table and the adaptive baselines of Section~\ref{sec:eval-adaptive}); what speculation gains over decoding without it is the drafter's, not the tuner's, and we show it only for context.

\subsection{Q1: measured tables against fixed lengths}
\label{sec:eval-tables}

\begin{table*}[t]
\centering
\small
\caption{Worst loss (\%) against the best fixed draft length at each load, over 1--128 open requests (Spec-Bench: 1--64), mean of three runs. \emph{Table} is \sys{}'s tuned table; \emph{Best fixed} is the one fixed draft length with the smallest worst-case loss (length in brackets), which an operator could only find by benchmarking every length. Lower is better; bold marks the better of the two. H100 V2: full CUDA graphs per draft length; H100 V1: vLLM's default runner, which switches to piecewise CUDA graphs whenever a table is set. Older GPUs: the same settings from our T4 and A10G study, each run on the GPU named (EAGLE-3 code on both). $^\dagger$single run. $^\ddagger$the \emph{0/1/3 table}, built from the tuner's length-0, 1 and 3 profiles only (the \emph{all-length table}, built from every profiled length, steps from length 2 to 1, which stalls this vLLM nightly; Section~\ref{sec:eval-engine}). $^\S$the tuned table is that single fixed length, so the two differ only by run-to-run noise. n/a: the tuner's V2 profile of the draft model is invalid (Section~\ref{sec:eval-cost}).}
\label{tab:main}
\vspace{3pt}
\setlength{\tabcolsep}{5pt}
\begin{tabular}{@{}l rr rr lrr@{}}
\toprule
 & \multicolumn{2}{c}{H100 V2} & \multicolumn{2}{c}{H100 V1} & \multicolumn{3}{c}{Older GPUs} \\
\cmidrule(lr){2-3}\cmidrule(lr){4-5}\cmidrule(l){6-8}
Setting & Table & Best fixed & Table & Best fixed & GPU & Table & Best fixed \\
\midrule
EAGLE-3, code & 4.0 & \textbf{3.4} (2) & 39.5 & \textbf{3.1} (2) & T4 / A10G & \textbf{4.0} / \textbf{2.3} & 9.6 (2) / 15.0 (1) \\
EAGLE-3, 4-bit AWQ target & \textbf{3.0}$^\ddagger$ & 4.6 (2) & 43.0 & \textbf{4.2} (2) & T4 & 20.6$^\dagger$ & \textbf{9.2} (1)$^\dagger$ \\
n-gram, code-apply & 5.7$^\S$ & 4.5 (5) & 5.2 & \textbf{0.0} (5) & T4 & \textbf{2.8} & 7.6 (5) \\
Qwen3-0.6B draft model, code & n/a & 0.0 (1) & 35.6 & \textbf{0.0} (1) & T4 & \textbf{3.5}$^\dagger$ & 12.2 (1)$^\dagger$ \\
DFlash, code & \textbf{11.6} & 19.1 (3) & 29.5 & \textbf{20.4} (3) & A10G & \textbf{5.2} & 18.0 (3) \\
EAGLE-3, Spec-Bench chat & 3.1$^\S$ & 3.9 (3) & 39.3 & \textbf{2.5} (2) & T4 & 3.8 & \textbf{2.2} (2) \\
\bottomrule
\end{tabular}
\vspace{-3mm}
\end{table*}

Table~\ref{tab:main} is the main result. For each GPU it compares the tuned table's worst loss with that of the one fixed length an operator would have to find by benchmarking every length.

\emph{On the H100 with V2}, the tuned tables stay within 3.0--5.7\% of the best fixed length at every load for EAGLE-3, the 4-bit target, n-gram and Spec-Bench, and within 11.6\% for DFlash, whose best single fixed length loses 19.1\%. Here a well-chosen single length is nearly as safe: 3.4\% for EAGLE-3, and for n-gram and Spec-Bench the tuner itself returns one length, the one an exhaustive benchmark picks (5 and 3). So on one H100 up to 128 requests the table's value is that it finds that length without benchmarking every length. Where no single length works, the table keeps finding the right one: beyond 128 requests, where in llm-d fixed length 3 falls up to 24\% behind it (Section~\ref{sec:eval-llmd}); on a mixture-of-experts target, with DFlash, and on the older GPUs (below).

\emph{A mixture-of-experts target} behaves differently. With Qwen3-30B-A3B (30B parameters, 3B active per token) and its EAGLE-3 drafter on one H100 with V2, the step without speculation grows from 4.7~ms at a batch of 1 to 19.7~ms at 128, because a larger batch activates more experts, while three draft tokens add 61\% at a batch of 1 but only 31\% at 128, once those experts are loaded. Speculation therefore keeps paying as load grows: length 3 is the fastest fixed length at every load from 2 to 96 requests and length 2 at 128 (by 2.3\%), and the tuner finds that table ($[[1,3,3],[4,4,2],[5,128,3]]$), within 2.9\% of the best fixed length at every load. A table written for a dense model does not fit here: vLLM's documented table switches to length 1 above 64 requests and loses 17.5\% at 96 and 12.2\% at 128 requests (three runs).

\emph{Long prompts} change the table less. With GovReport prompts~\cite{huang2021govreport} of 1{,}024 and 3{,}072 tokens (EAGLE-3, three runs), the re-profiled table moves its edge from a batch of 50 to 36--38 and stays within 0.7\% and 1.5\% of the best fixed length, but the 32-token table (2.0\%) and vLLM's documented table (2.3\%) do nearly as well: on the H100 the step grows slowly enough that prompt length barely matters.

\emph{On the older GPUs}, for full-precision targets on the code workloads, the tables stay within 2.3--5.2\% where the best single fixed length loses 7.6--18.0\%, and every fixed length loses 7.6--43.6\% at some load.

\emph{On the H100 with the default V1 runner}, the same kind of table loses 30--43\% for every model-based drafter (n-gram, which adds no model pass, 5.2\%): the cost of the engine's graph mode, not of the tuner's choices (Section~\ref{sec:eval-engine}).

\paragraph{Jumping load.} We drove one server per configuration through eleven 50-second windows whose open requests jump between 2 and 128 (three runs). On the A10G the tuned table never fell below 81\% of the best configuration in a window (mean 94.0\%). On the H100 with V2 it was the best configuration in practically every window: \textbf{98.5--100\%} of the per-window best (mean 99.6\%), 4.1\% above fixed length 1, 1.1\% above the documented table and level with fixed length 2 ($-$0.2\%). With V1, fixed length 2 beat it by 15\%.

\paragraph{Open-loop arrivals.} We also sent requests at 2 to 72 per second with Poisson and with bursty arrivals (gamma inter-arrival times, smoothness 0.25), on the H100 with V2, three runs each. Below saturation the draft length shows in latency, and no single fixed length is right at every rate: at 2 requests per second the median request takes 0.63--0.64~s with the tuned table and fixed length 3 and 18\% longer with length 1; at 72 per second the table takes 1.08~s, length 1 1.10~s and length 3 22--24\% longer. Bursty arrivals change none of these orderings.

\subsection{Q2: against prior adaptive control}
\label{sec:eval-adaptive}

\begin{figure*}[t]
\centering
\includegraphics[width=\textwidth]{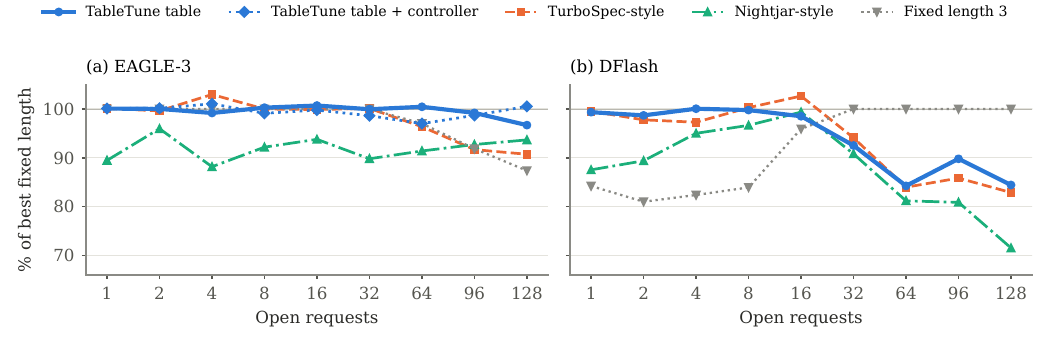}
\vspace{-6mm}
\caption{\sys{} against TurboSpec-style and Nightjar-style control on the H100 (V2 runner), code prompts at temperature 1.0, mean of three runs. Throughput as \% of the best fixed draft length at each load; every method chooses among the lengths of its starting table: with EAGLE-3, 0/1/3 for TurboSpec- and Nightjar-style, 2/3 for the all-length tuned table and 1/3 for the 0/1/3 table (also the controller's start); with DFlash, 2/4/6/8 for all.}
\label{fig:baselines}
\vspace{-3mm}
\end{figure*}

\paragraph{Baselines.} TurboSpec and Nightjar released no code, so we reimplemented each as a vLLM plugin following its paper, in its own package that shares no code with \sys{}. \emph{TurboSpec-style} picks the length every step, for the whole batch, maximizing goodput from a linear step-time model (a target pass over $B(k{+}1)$ tokens plus $k$ drafting passes) fitted to the tuner's profile at eight batch sizes, and from one per-token acceptance rate, a moving average over steps, with the geometric formula for tokens per step~\cite{liu2024turbospec}. \emph{Nightjar-style} runs Nightjar's ADA-BINGREEDY bandit for every batch size, with the measured time per output token as its loss~\cite{li2025nightjar}; offloading the drafter to CPU memory needs engine changes and is omitted. Both start from the tuner's acceptance profile where they use one, and, like \sys{}'s controller, may only choose lengths that are in their starting table, the only ones with full CUDA graphs under V2.

\paragraph{Results.} With EAGLE-3 (Figure~\ref{fig:baselines}a) the tuned tables stay within 3.3\% (all-length table) and 3.7\% (0/1/3 table) of the best fixed length at every load, 3.0\% with the controller, TurboSpec-style within 9.3\% and Nightjar-style within 11.8\%. TurboSpec-style matches the tables up to 32 requests and then keeps length 3 too long: its linear model ranks length 3 first up to a batch of about 100, while the measured profile switches to length 2 above 50, because the fit smooths over the jump in length-3 step time (at a batch of 80, measured 9.7~ms, fitted 8.0~ms). The same choice makes its requests 20--23\% slower than with the table at 72 open-loop requests per second. Nightjar-style loses 4--12\% at every load because each batch size learns separately and keeps exploring. With DFlash (Figure~\ref{fig:baselines}b) all three trail fixed length 3 by 10--29\% above 32 requests, a length none of them has: the tuned table uses 2/4/6/8 and predicted length 4 to beat 3 there, which the measured sweep contradicts (Section~\ref{sec:eval-cost} explains why); among the three, \sys{} loses least (15.8\%, against 17.1\% and 28.5\%). In a repeat of the jumping-load study that adds the two baselines (greedy, three runs), the tuned tables never fall below 95\% of the best configuration in a window (the all-length table: 98\%; 98.8--99.2\% on average); TurboSpec-style and Nightjar-style fall to 89\% and 92\% in the high-load windows and deliver 5.3\% and 4.1\% fewer tokens in total than the all-length table. With greedy acceptance fixed length 2 is nearly as good as the tables here (98.3\% on average).

\paragraph{vLLM's adaptive verification.} vLLM's \emph{adaptive verification}~\cite{vllm2026adaptiveverif,vllm2026avnohead} picks at every step how many drafted tokens of each request to verify, from a startup cost profile and acceptance predicted from the draft logits; it needs the V2 runner and variable-length attention, and does not support n-gram drafting, LoRA or pipeline parallelism. On vLLM main of October 20, 2026 (EAGLE-3, 1--128 requests, three runs), the tuner, rerun in 6 minutes, gives $[[1,36,3],[37,128,2]]$, within 3.4\% of the best fixed length at every load (3.1\% with the controller; fixed length 2: 4.1\%). Adaptive verification with up to 3 tokens trails the best fixed length by up to 26.2\% and fixed length 3 by 7--26\% at 2--64 requests, accepting 0.95 tokens per draft against 1.11. It was designed for confidence-head drafters on large mixture-of-experts models, so this one-model result is preliminary.

\subsection{Q3: the controller}
\label{sec:eval-ctrl}

\begin{figure*}[t]
\centering
\includegraphics[width=\textwidth]{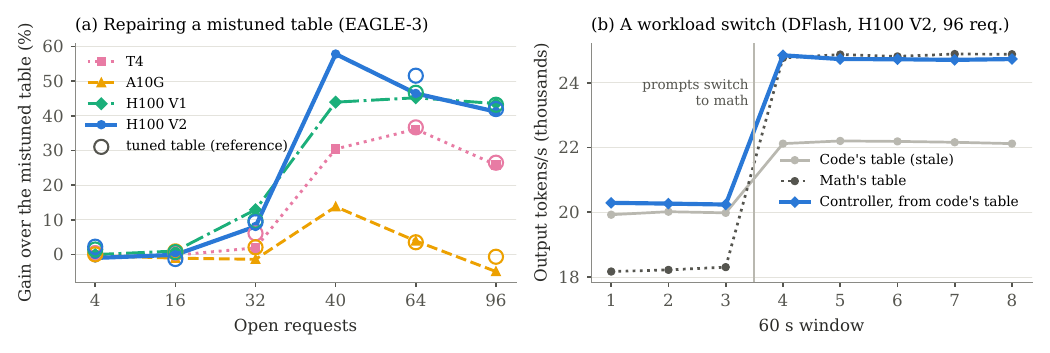}
\vspace{-6mm}
\caption{The controller. (a) Repairing the same mistuned table on three GPUs: vLLM's documented table with its ranges divided by 4 (length 0 above a batch of 32), started with a mismatched chat prior, EAGLE-3 on code. Lines: gain of the controller over the same table without it; hollow circles: gain of that GPU's tuned table, the reference the controller should reach (no sweep point at 40). (b) A workload switch: DFlash on the H100 (V2), 96 open requests, code prompts, then math from the fourth 60-second window. Mean of three runs.}
\label{fig:ctrl}
\vspace{-3mm}
\end{figure*}

\begin{table*}[t]
\centering
\small
\caption{The controller: best gain over the same table without it, at the load where it occurs (loads 4--96; n-gram 4--128), mean of three runs unless marked ($^\dagger$single run). Top: steady traffic, starting from a mistuned table. Bottom: a mid-run workload switch (Spec-Bench math and translation with EAGLE-3; DFlash: code then math at 96 requests; greedy, A10G protocol), the stale table's loss and what the controller with exploration and prior decay does about it. Older GPUs: the same experiment from our T4 and A10G study, on the GPU named. $^\S$still 60\% below no speculation, which beats every draft length of this drafter under V2.}
\label{tab:ctrl}
\vspace{3pt}
\setlength{\tabcolsep}{6pt}
\begin{tabular}{@{}llll@{}}
\toprule
Starting table & H100 V2 & H100 V1 & Older GPUs \\
\midrule
Documented table / 4 (Figure~\ref{fig:ctrl}a) & \textbf{+58\%} (40) & +45\% (64) & T4 +36\% (64), A10G +14\% (40) \\
Documented table & +2\% (16) & +8\% (96) & T4 +15\% (64), A10G +16\% (64) \\
n-gram, hand-written table & \textbf{+30\%} (4)$^\dagger$ & +22\% (4)$^\dagger$ & T4 +24\% (16) \\
Draft model, EAGLE-3's table & +40\% (96)$^{\dagger\S}$ & +2\%$^\dagger$ & T4 +1\%$^\dagger$ \\
DFlash, EAGLE-3's table & +23\% (64) & +37\% (64) & A10G +11\% (4) \\
4-bit AWQ, documented table & +5\% (64)$^\dagger$ & $-$9 to $-$24\%$^\dagger$ & T4 +23\% (64)$^\dagger$ \\
\midrule
Workload switch: stale table's loss & $-$4.5 to 0.4\%; DFlash 10.8\% & 1.0--1.5\% & A10G 7.5--18.1\% \\
\quad controller (both options) & $-$4.4 to +1.4\%; DFlash recovers 96\% & $-$2.5 to +3.2\% & A10G recovers 72--85\% \\
\bottomrule
\end{tabular}
\vspace{-3mm}
\end{table*}

\paragraph{Repairing mistuned tables.} A table is \term{mistuned} when its ranges or lengths do not fit the GPU, drafter or workload it serves. The most natural case is vLLM's documented table, deployed as an operator copies it. In llm-d with Qwen3-32B it drops to length 0 above a batch of 128 per replica, and started from it the controller raises throughput by 9--29\% at 256--1024 open requests, to the tuned table's level (Section~\ref{sec:eval-llmd}); on one H100 up to 128 requests the same table is nearly right, and the controller adds only 2\%. To test repair under larger errors, Table~\ref{tab:ctrl} and Figure~\ref{fig:ctrl} also start it from deliberately mistuned tables, with an acceptance prior measured on a different workload. From the documented table with its ranges divided by 4, which drops to length 0 above a batch of 32, it gains up to 58\% on the H100 (V2) and 45\% (V1), against 36\% on the T4 and 14\% on the A10G: the H100's wider gap between speculation and none makes a mistuned table costlier and its repair larger. At every load and on every GPU it reaches 96--103\% of the tuned table's throughput (Figure~\ref{fig:ctrl}a, circles). On the mixture-of-experts target the same mistuned table costs far more, because speculation pays at every load there: the controller raises throughput by 71--83\% at 40--96 requests (three runs), again to the tuned table's level. Started from another drafter's table it rebuilds the right one: +23\% for DFlash on the H100 with V2, reaching the tuned table's throughput with only the starting table's lengths 1--3 (V2 has full graphs only for those), and +37\% under V1, where it may use any profiled length. At loads where the starting table is already near-optimal it changes throughput by $-$3.7 to +3\%. Under V1 on the H100's 4-bit target it costs 9--24\% (single run, a launch-bound step); under V2 the cost disappears.

\paragraph{A workload switch.} We switched the prompts mid-run between Spec-Bench math and translation, the pair whose tables differ most, at the load where they disagree (80 open requests on the A10G and the H100 with V2, 71 with V1); after the switch the first workload's table is \term{stale}. On the A10G, where translation's table runs length 0 and math's length 2 there, the stale table loses 7.5--18.1\%; the controller with its defaults oscillates or stays at length 0, where it observes nothing, while with exploration and prior decay it recovers 72\% and 85\% of the gap and rebuilds math's table exactly. We therefore recommend both options as defaults. On the H100 the same EAGLE-3 switch leaves little to recover ($-$4.5 to 1.5\%, Table~\ref{tab:ctrl}): EAGLE-3's acceptance varies too little across workloads for its H100 tables to differ much. DFlash's varies far more (5.03 tokens per step at length 8 on math, 2.51 on translation), so we repeated the switch with DFlash, code then math at 96 open requests, where code's table runs length 2 and math's length 6 (Figure~\ref{fig:ctrl}b). The stale code table loses 10.8\% after the switch (three runs); the controller rebuilds math's table within the first minute and recovers 96\% of the gap (+11.7\%, within 0.4\% of math's own table), and before the switch it beats code's own table by 1.5\% by learning that length 4 pays at this load. TurboSpec-style also recovers it (+12.4\%), since it too tracks acceptance online; Nightjar-style recovers 38\%, because its bandit has to relearn every batch size. From translation to math at 48 requests the tables differ (length 4 against 8) but their throughput does not ($-$1.1\%), and every controller stays within 0.6\% of math's own table.

\subsection{Q4: end to end in llm-d}
\label{sec:eval-llmd}

\begin{figure*}[t]
\centering
\includegraphics[width=\textwidth]{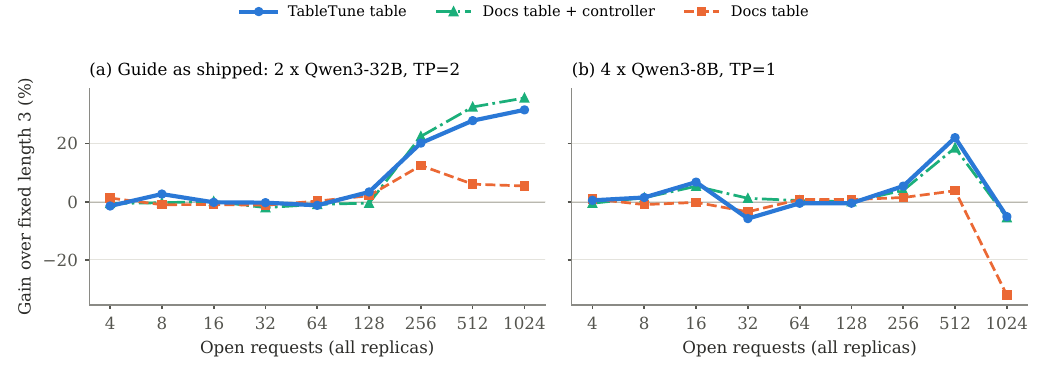}
\vspace{-6mm}
\caption{llm-d optimized-baseline with EAGLE-3 behind the llm-d router, code prompts at temperature 1.0, mean of three runs: output tokens/s against fixed draft length 3 (0 = fixed length 3). (a) The guide's own model and layout, 2 $\times$ Qwen3-32B at TP=2 on four H100s. (b) 4 $\times$ Qwen3-8B at TP=1 on the same GPUs; the 32-request dip is shared by every speculative configuration. The tuned table with the controller is within 2\% of the table alone up to 512 requests in (b), and up to 4\% above it in (a); it is omitted.}
\label{fig:llmd}
\vspace{-3mm}
\end{figure*}

\paragraph{Setup.} llm-d's optimized-baseline guide~\cite{llmd} serves vLLM replicas behind the llm-d router, an Envoy proxy with an endpoint picker that routes by prefix-cache affinity and token load. We deployed the guide as shipped, two replicas of Qwen3-32B at TP=2 on four H100s with the guide's Helm values for the router, which are calibrated for this setup, and changed only configuration: a Kustomize overlay on the guide's model-server base adds the model's EAGLE-3 drafter and the tuned table to \texttt{--speculative-config}, uses a vLLM nightly image with the V2 runner, caps each replica at 256 sequences of 4096 tokens, and for the controller puts the \sys{} package on \texttt{PYTHONPATH} and sets its \texttt{VLLM\_DSD\_*} variables; the overlay is in the repository. The tuner profiled one replica at TP=2 with the same image in about 30 minutes; its all-length table steps from length 2 to 1, which stalls this nightly (Section~\ref{sec:eval-engine}), so we deploy its 0/1/3 table, $[[1,60,3],[61,256,1]]$. As a second scale we ran four Qwen3-8B replicas at TP=1 on the same GPUs. AIPerf sent code prompts through the router at 4--1024 open requests, three runs per configuration (run-to-run spread 3.2\% median).

\paragraph{Results.} On the guide as shipped (Figure~\ref{fig:llmd}a), up to 128 open requests the tuned table runs length 3 and matches fixed length 3 and the documented table within run-to-run noise. From 256 the replicas cross the table's edge to length 1, and there it is 20--32\% faster than fixed length 3 and 7--25\% faster than the documented table, which drops to length 0 above a batch of 128 per replica; at 512 and 1024 open requests the median time between tokens is 23--29\% lower than with either, and the median first token arrives 23--26\% sooner at 1024. (Speculation itself, with any of these configurations, adds 48--78\% over the guide as shipped at up to 64 requests; that gain belongs to the drafter.) Started from the documented table, the controller raised throughput by 9--29\% at 256--1024 open requests, to the tuned table's level, and on top of the tuned table it added up to 4\%. With four Qwen3-8B replicas (Figure~\ref{fig:llmd}b) the table is up to 22\% faster than fixed length 3 (5\% behind it at 1024) and up to 40\% faster than the documented table, and the controller adds 14\% and 39\% to the latter at 512 and 1024 requests.

\subsection{Q5: what the engine must provide}
\label{sec:eval-engine}

\paragraph{Full CUDA graphs per draft length.} On vLLM's default V1 runner any table switches the decode step to piecewise CUDA graphs, which add CPU launch work to every step (a Nsight trace on the T4 shows the GPU busy 91\% of the time with full graphs and 60\% with piecewise ones). A long step hides that work, as on the older GPUs; the H100's step is short (about 6~ms for Qwen3-4B), so it shows: piecewise graphs alone cost fixed length 3 24--34\% at 1--32 requests, more host cores do not help (2, 4 or 16), and host contention makes it worse. With V2's full graphs per draft length the loss disappears (Table~\ref{tab:main}). Until an engine provides them, keep a table only if it beats the best fixed length and no speculation, both timed with full graphs.

\paragraph{Length switches that do not stall.} On this vLLM nightly, several tables that step from length 2 to length 1 stalled the V2 runner under load (the 4-bit target's tuned table, EAGLE-3's T4 table, translation's table, and llm-d's at its 80/81 edge). The 0/1/3 tables, and tables without such a step, never stalled; we report those where marked, and the fault upstream.

\paragraph{A real off switch.} A table entry of 0 still runs EAGLE-style drafters, which keep their own KV cache: where no speculation is best (the A10G above 96 requests) the tuned table trails it by 5.6--6.7\%, and on the H100 a table at length 0 runs 19--22\% below no speculation. Nightjar~\cite{li2025nightjar} provides one.

\subsection{Q6: cost and reliability of measuring}
\label{sec:eval-cost}

\paragraph{Overheads.} Profiling takes 8 minutes on the H100 (20 for the mixture-of-experts target, 28 for Qwen3-32B at TP=2, 22 on the A10G, 45 on the T4), once per GPU, engine, model and drafter; a new workload needs only a 2-minute acceptance profile. The controller adds no GPU work.

\paragraph{Measure, do not fit.} Step times must be measured per GPU and engine. On the H100 a linear model misses a jump in length-3 step time and costs TurboSpec-style control 9\% (Section~\ref{sec:eval-adaptive}); on the A10G every simple model we fitted misses the switch to length 0 and loses 12.7\% at 128 requests.

\paragraph{Where the profile misses: DFlash.} Timing with every guess rejected does not time work that grows with accepted tokens. For DFlash that matters: its acceptance in the sweep matches the profile, but at length 8 its steps take 1.3--1.9$\times$ the profiled time at live batches of 6--38, against 1.05--1.16$\times$ at length 1. Profiling with the measured acceptance instead (vLLM's synthetic rejection sampler, steps counted from its draft counter) corrects the step times but barely changes the table: in both profiles length 3 wins only in short ranges above a batch of about 47, which the 25\% minimum-width rule merges into length 4. A tolerance repairs this: merge a narrow range only if that loses less than 2\% of its predicted goodput, our tie threshold. It changes four of our V2 tables; rerun against the original ones in the same session (three runs), it makes DFlash 6.5\% faster at 64 requests, n-gram about 5\% faster at 96--128 and Spec-Bench 3\% faster at 64, and leaves EAGLE-3 within noise; the other tables do not change. It is a tuner option (\texttt{--merge-tol}); the results above use the original rule. DFlash still trails fixed length 3 by 10\% at 128 requests, serving work such as interleaved prefill that a decode-only profile cannot see.

\paragraph{Near-ties and an invalid profile.} With EAGLE-3 on the H100, lengths 2 and 3 nearly tie, and two profiles of the same GPU differed by 4--9\% in single step times, enough to move an edge at a cost of a few percent. Under V2 the profile of the Qwen3-0.6B draft model is flat at about 40~ms, against 9--37~ms when served, so that table is invalid.

\section{Limitations}
\label{sec:limits}

\paragraph{Engines.} The results with full graphs use one vLLM nightly, whose V2 runner stalls when a table steps from length 2 to 1 (Section~\ref{sec:eval-engine}); where that matters, including both llm-d deployments, we report the 0/1/3 table, which can be slightly worse than the all-length table.

\paragraph{Accuracy of the table.} The tuner's choices are only as good as its profiles. With DFlash the table loses up to 16\% to fixed length 3 at batches of 34--87; a merge tolerance repairs part of it, but serving work the profile cannot see leaves a 10\% gap at 128 requests (Section~\ref{sec:eval-cost}), and the controller, whose prior was accurate, does not correct cost errors. The draft model's V2 profile was invalid, and its T4 results at high load are decided by KV-cache pressure and preemption, which the step-time profile cannot see.

\paragraph{Scope.} Dense targets of 4B, 8B and 32B parameters and one mixture-of-experts target with 3B active parameters, on H100s in a single node at TP of at most 2; prompts of 32 tokens in most studies and of up to 3{,}072 tokens in one, answers up to 256 tokens (512 for n-gram); closed-loop load, plus open-loop arrivals up to 72 requests per second on one GPU. Prompts beyond 4{,}096 tokens, mixed and recorded traffic, larger mixture-of-experts models, multi-node and disaggregated serving are missing. TurboSpec and Nightjar released no code; our reimplementations omit Nightjar's drafter offloading and TurboSpec's context-length term. We compare runs only within a node and session.

\section{Related Work}
\label{sec:related}

\paragraph{Drafters and per-request adaptation.} Drafter designs~\cite{li2024eagle,li2024eagle2,li2025eagle3,cai2024medusa,miao2024specinfer,chen2024sequoia,saxena2023prompt,fu2024lookahead,chen2026dflash} and methods that adapt the length per request or token~\cite{mamou2024disco,huang2025specdecpp,agrawal2024adaedl,hou2025banditspec,liu2025pearl,zhang2025svip} or update the drafter online~\cite{liu2024online} are complementary to choosing the length from load.

\paragraph{Load-aware speculation in serving.} Su et al.~\cite{su2023synergy} proposed choosing the speculation length per batch size, the idea behind vLLM's Dynamic SD table. BASS~\cite{qian2024bass} batches speculation, TETRIS~\cite{wu2025tetris} picks which draft tokens to verify across a batch, and Sequoia~\cite{chen2024sequoia} sizes its token tree for the hardware. TurboSpec~\cite{liu2024turbospec} is closest to our work: it maximizes goodput inside vLLM with a linear step-time model fitted to profiled runs and a moving average of recent acceptance, choosing the length at each step. \sys{} applies these ideas to vLLM's production table, measures the grid instead of fitting it, and keeps the table current with a plugin; Section~\ref{sec:eval-adaptive} compares it with reimplementations of TurboSpec and Nightjar. Nightjar~\cite{li2025nightjar} picks the length per batch size online with a multi-armed bandit, switches speculation truly off and moves the drafter to CPU memory. vLLM's adaptive verification~\cite{vllm2026adaptiveverif,vllm2026avnohead} sizes the verified drafts per request and step from a profiled cost model and predicted acceptance (Section~\ref{sec:eval-adaptive}). SpecServe/AdaSpec~\cite{huang2025adaspec} and AdaServe~\cite{li2026adaserve} adapt speculation to latency targets; we optimize throughput. MagicDec~\cite{sadhukhan2025magicdec} finds speculation pays at large batches for long contexts, and Kong et al.~\cite{kong2026latency} model its latency under load.

\section{Conclusion}

A draft-length table must be measured for each GPU, engine and model and kept current as traffic changes; \sys{} measures it in minutes and repairs it while serving. On an H100 with full CUDA graphs its tables stay within 3--6\% of the best fixed length for most drafters, with lower worst-case loss than reimplemented TurboSpec- and Nightjar-style control and vLLM's adaptive verification. Its controller recovers up to 58\% from mistuned tables, and in llm-d's optimized-baseline guide as shipped the tuned table is up to 32\% faster than a fixed draft length and 25\% faster than vLLM's documented table.

Tables do not transfer (a dense-model table loses up to 18\% on a mixture-of-experts model); engines need full graphs per draft length, stall-free switches and a real off switch.

\bibliography{paper}
\bibliographystyle{mlsys2025}

\section*{AI Usage Disclosure}

During the preparation of this work, Anthropic's Claude Sonnet was used to assist the authors in summarizing preliminary literature during the background phase and structuring the initial outline of the related work section. The tool was also used to generate boilerplate Python scripts for rendering the graphs. Anthropic's Claude Code was used for submission formatting checks. The authors manually scrutinized all generated scripts, conducted the actual evaluations, and independently drafted the final text. The authors take full responsibility for the manuscript's accuracy and integrity.

\end{document}